\documentclass[11pt,a4paper]{article}
\pdfoutput=1

\usepackage[english]{babel}
\usepackage{epsfig}

\usepackage[utf8]{inputenc}
\usepackage{floatrow}
\usepackage{amssymb,amsmath,amsfonts,amsthm,graphicx,psfrag}
\usepackage{indentfirst}
\usepackage{hyperref}

\usepackage[title,titletoc]{appendix}
\usepackage{graphicx}
\usepackage{amsfonts}
\usepackage{bm}
\usepackage{verbatim}
\usepackage{epsfig}

\graphicspath{{./pictures/}}

\usepackage{mathrsfs}
\usepackage{amsthm}
\usepackage{amssymb}
\usepackage[T1]{fontenc}

\usepackage{authblk}
\usepackage{cite}
\usepackage{slashed}
\usepackage{amsmath}
\usepackage[left=2cm,right=2cm,
    top=2cm,bottom=2cm,bindingoffset=0cm]{geometry}
\newcommand{\be}{\begin{eqnarray}}
\newcommand{\ee}{\end{eqnarray}}

\def\ga{\mathrel{\mathpalette\fun >}}
\def\fun#1#2{\lower3.6pt\vbox{\baselineskip0pt\lineskip.9pt
\ialign{$\mathsurround=0pt#1\hfil ##\hfil$\crcr#2\crcr\sim\crcr}}}

\newcommand{\vex}{\mbox{\boldmath${\rm x}$}}
\newcommand{\vey}{\mbox{\boldmath${\rm y}$}}

\newcommand{\lan}{\langle}
\newcommand{\ran}{\rangle}

\usepackage{authblk}

\title{Speed of sound, breaking of conformal limit and instabilities in quark-gluon plasma at finite baryon density.}
\author[*,+]{Z.V Khaidukov}
\author[*]{Yu.A.Simonov}
\affil[+]{Moscow Institute of Physics and Technology,
Institutskiy per. 9, 141700 Dolgoprudny, Moscow Region, Russia}
\affil[*]{Institute for Theoretical and Experimental Physics, B. Cheremushkinskaya 25, Moscow, 117259, Russia}

\begin{document}
  \maketitle
\begin{abstract}

The properties of the quark-gluon plasma(QGP) in the presence of baryon chemical potential are studied
using the Field Correlator Method(FCM). At low densities the QGP thermodynamics with the colormagnetic confinement 
and the Polyakov line interaction is in good agreement with lattice 
data,and the speed of sound satisfies $C_s^2\leq \frac13$,but in the intermediate range
of densities the speed of sound displays a singular behaviour and one needs 
density modifications of the Polyakov line interaction to avoid 
instabilities.   In particular  for $\mu_q >0.5$ GeV there exists a region in the $(\mu, T)$ plane where $C^2_s$ violates  the  conformal  limit. This behaviour is strongly connected to the properties of the  Polyakov loop and its  density  dependence.
\end{abstract}

\section{Introduction}

The main result of heavy ion experiments performed over the last 15 years at RHIC and then at RHIC and LHC  is the discovery of a new form of matter \cite{QGP1,QGP2,QGP3,QGP4,QGP5} with its properties markedly different from the pre-RHIC era predictions-see \cite{HIC1,HIC2,HIC3,HIC4,HIC5,HIC6,HIC7,HIC8,HIC9,HIC10} and references therein. Instead of the commonly assumed picture of weakly coupled Quark-Gluon Plasma(QGP) a strongly coupled liquid has emerged,  subject  to the  law of   the relativistic  hydrodynamics\cite{Hydro1,Hydro2,Rev}. The properties  of the produced matter are drastically  changing as it passes several stages  of evolution: from formation,   hydrodynamization and thermalization toward the hadron gas   production. The wealth of the QCD matter phases is reflected in the QCD phase diagram drawn in the ($\mu,T$) plane. However, the correspondence between the specific($\mu,T$) domains of the phase diagram and the space-time dynamics of the fireball should be considered with caution.
The reason is that the phase diagram describes the limit of  an infinite system in thermodynamic equilibrium.\par
On the theoretical side the matter created in heavy ion collisions should be described by the fundamental laws of QCD.  
 The  dynamics and thermodynamics of the Quark-Gluon Plasma (QGP)  is now in the focus of numerous investigations \cite{Invi}. The presence of strong interaction in QGP at zero baryon density was demonstrated in  numerous lattice data \cite{qlatt1,qlatt2,
 latt1,latt3,latt4,lattglu4}, which show that the ratio of the QGP pressure to the Stefan-Boltzmann value $P_{SB}$ is around 0.8 and  remains almost constant up to 1 GeV, implying a strong interaction  growing  with $T$.

Another striking discovery was the analysis of the temperature transition, made in the $2+1$ QCD lattice  computations, which showed a smooth crossover in the temperature region $T=140\div 180$ MeV \cite{25*}\footnote{This QCD crossover  is a new phenomenon, possibly having some analogs in the material sciences  and in the ionization  and  dissociation  processes.} But the question of the existence of a critical point at  finite baryon  chemical potential is still of intense interest.\cite{Crit}

As a result  the question about the  structure  of the QCD  phase diagram remains open on the  lattice side.
This  happens mostly  because lattice methods  are  strongly restricted   to a  domain of small chemical potentials ($N_c$=3)  due to the "sign problem".  To circumvent this difficulty  in the case of  $N_{c}=3$    one   exploits  the  Taylor  expansion  around zero chemical potential  \cite{thLatt1,thLatt2}, or    imaginary  chemical potential\cite{thLatt3}. Another possibility is to use on the lattice  the number of colors $N_c=2$, where the sign problem is absent \cite{Kot1,Kot2,Kot3,Kot4}.

On the theoretical side one can exploit the method, which is applicable at any chemical potential and any temperature -- the Field Correlator Method (FCM), where the   nonperturbative  dynamics in the confinement and deconfinement regions is based on vacuum properties, described by gluonic field correlators \cite{VCM2,VCM3,VCM4,VCM5,VCM6,VCM7}.

The strength of this method  is connected with  the possibility of a complete self-consistent description of both  QGP plasma (also in the presence of the chemical potential) and  the  hadronic matter in the confinement phase  \cite{3,19,20,21,22,23,24,25,26}. The main  ingredient  is the vacuum  average  of colorelectric fields $D^E$ and colormagnetic  fields $D^H$, which provide colorelectric confinement (CEC)  with the string tension $\sigma (T)$ and colormagnetic  confinement (CMC) with the string  tension $\sigma_s (T)$.  
The latter,   calculated   from field  correlators and  on the  lattice grows with $T$ , $\sigma_s (T) \sim g^4 (T) T^2$, and insures the strong interaction at large $T$, mentioned above.

From the point of view of FCM the crossover phenomenon     is  connected with  the gradual vanishing of the   vacuum confining correlator $D^E(z)$ (and the resulting string tension $\sigma (T)$) with  the growing temperature. The same phenomenon of the ``melting  confinement'' can be observed in the SU(3) gluondynamics \cite{23},  where also the decreasing with $T$ string tension $\sigma(T)$, measured on the lattice \cite{45,46,47,48}, explains the behaviour of pressure for $T<T_c$, but in the case of SU(3) it cannot smoothly match the fast growing gluon pressure (in contrast to the slowly  growing glueball pressure due to  large glueball masses $\ga 2$ GeV). As a result, one has in SU(3) a weak  first order transition,  \cite{23}
while in the $n_f = 2+1$ QCD with low mass  mesons  the vanishing of $\sigma (T)$ is complete in the course of transition.\par

As a proof of this picture one has the vanishing with $T$ the quark condensate \cite{Baz} which is  connected in FCM to the  confinement  $\lan \bar qq(T)\ran \sim \sigma^{3/2} (T)$ \cite {49,50,51}.\par

 Hence the FCM picture of the temperature transition is a vacuum based process, different from standard theoretical models, and this crossover does not imply the presence of a critical point. \par

The  first  study  of the $QGP$ thermodynamics at nonzero baryon  density using FCM,  and CMC was done in \cite{25}. In general, the main interaction in QGP  is provided by  the colormagnetic confinement, operating both below and above transition temperature, as was observed in lattice data \cite{42*}, where the CMC correlators $\lan tr F_i (x) \phi (x,y) F_{ik} (y)\ran$ have been measured.

It was found in \cite{42**} that CMC does not support white bound states in $q\bar q$ and $gg$  systems, however it can create the screening mass $M(T)$ of isolated quarks and gluons \cite{22,23,24,25,42***}, which grows with temperature, so that the ratio $\frac{M(T)}{T}$ is constant up to the logarithmic terms. 

As was shown in \cite{25}, CMC mechanism produces the  square root singularities in $P(\mu,T)$ in the  complex $\mu$ plane, which are  situated at the distance $\frac{i\pi}{T}$ over the real axis. But  the pressure still  can be calculated for  however large values of the chemical potential (for example in \cite{25} the pressure is calculated for $\mu=400$ MeV that corresponds   to the baryon chemical potential  $m_{B}\simeq 3\mu=1200$ MeV) and  it does not show any singularities.
However  the  study at higher $\mu$ done in the present paper, discovers new interesting phenomena.

We shall show in this paper that the behavior of QGP using CMC and the 
mu-independent Polyakov line shows a good agreement with lattice data at 
low densities,$\mu_{q}<\mu_{crit}=400$ MeV,but at larger mu and T=1.3 $T_{c}$ the sound
velocity strongly violates the conformal limit and becomes a singular 
function of mu. 
\par

It is  created by the strong NP dynamics, which gives large values  to $\frac{\partial^{2}P}{\partial T^{2}}$  and        $\frac{\partial^{2}P}{\partial T \partial\mu}$;   supported by    high values of $\frac{d^{2}L}{dT^{2}}$, where $L=exp(-F/T)$ is the Polyakov line value. 
\par
As will be shown below in the paper,to avoid this problem one should 
take into account the density modification of the Polyakov line.We shall use
in what follows the speed of sound as an indicator of the instability 
effects in QGP.

\par    

We need to point out that the question of the possible values of the speed of sound is of interest in itself. It plays a  very important role in the  physics of the medium   \cite{bookHen}.
The  question about  limitations of the speed of sound  at finite density  QCD is still open, because  at intermediate densities  the  standard (perturbative) QCD  theory is unable to predict  the value of $C^{2}_{s}$, but    in some models like ADS/CFT,   the upper limit 1/3 was found \cite{speed2,Sakai,D3D7,speedN2,speedholo}. A counterexample  was demonstrated  by Zeldovich \cite{speed11}(he used the mean-field and quasiclassical  approximation but the validity of these results at high density is questionable) and by Son (the medium with isospin chemical potential)  \cite{Son},and even in case of ADS/CFT correspondence there are a few counterexamples \cite{contrHolo1,contrHolo2}
Important restrictions could be found from the physics of neutron  stars.
For example,  the possible maximal mass exceeding two solar masses  might require the EoS with $C^{2}_{s}>1/3$ \cite{speed1,Blinnikov}.
\par
 Below  in this paper we   find  domains, where  the speed of sound  exceeds $C^{2}_{s}\ge 1/3$, for the used values of $L(T, \mu)$.

 \par The paper is organized as follows.  In section 2 we introduce the FCM in case of finite temperatures and densities. In section 3 we discuss the definition of the  speed of sound at finite density, and make some theoretical predictions, which we confirm in     section 4  using numerical results.   In section 5 we  discuss the behaviour of the Polyakov line at finite baryon densities, and will show that there is exist a renormalization of the Polyakov line that  leads to  a good behaviour of the system, and show,that there exists a density renormalization of the Polyakov
line which keeps the conformal limit unbroken.. The section 6  - final discussions and conclusions. \par

\section{The Field Correlator Method}
The FCM  is a useful instrument to treat the physics outside the area of perturbative theory.   Analysis of  physics of QGP  in terms of FCM   made in \cite{3,19,20,21,      
TVCM1,TVCM2}, has shown  the important role of Polyakov loops for description of thermodynamic of QGP, while in  \cite {22,23,24,25,26} also the CMC interaction was taken into account, providing a selfconsistent dynamical picture in a good agreement with lattice data. In the FCM the basic interaction of a quark or a gluon can be expressed via world lines affected by the
vacuum fields and finally written in the form of Wilson loops and Polyakov lines. It is essential that in the deconfined phase two basic interactions
define  quark  and gluon  dynamics: the colorelectric (CE) interaction, contained in the Polyakov
line L(T), and the colormagnetic (CM) one in the spatial projection on the Wilson loop.
 In the FCM the string tension is defined as: 
\begin{eqnarray}
\sigma^{E,H}=\frac{1}{2}\int{D^{E,H}d^{2}z}
\end{eqnarray}
where $D^{E,H}$  is obtained from
\begin{eqnarray}
\frac{g^{2}}{N_{c}}\ll Tr E_{i}(x)\Phi E_{j}(y)\Phi^{+} \gg=\delta_{ij}(D^{E}(u)+D^{E}_{1}(u)+u^{2}_{4}\frac{\partial D^{E}_{1}(u)}{\partial u^{2}})+u_{i}u_{j}\frac{\partial^{2} D^{E}_{1}(u)}{\partial u^{2}} \\
\frac{g^{2}}{N_{c}}\ll Tr H_{i}(x)\Phi H_{j}(y)\Phi^{+} \gg=\delta_{ij}(D^{H}(u)+D^{H}_{1}(u)+\textbf{u}^{2}\frac{\partial D^{H}_{1}(u)}{\partial \textbf{u}^{2}})-u_{i}u_{j}\frac{\partial^{2} D^{H}_{1}(u)}{\partial u^{2}}, 
\end{eqnarray} 
where $u = x-y$ and $\Phi(x,y)=Pexp(\int^{x}_{y}{A_{\mu}dz^{\mu}})$.

Using the T dependent path integral (world line) formalism one can express thermodynamic
potentials via the Wilson loop integral, e.g. for the gluon pressure one has \cite{3,21,23}
 \be
 P_{gl}=2(N^{2}_{c}-1)\int_{0}^{\infty}\frac{ds}{s}\sum_{n=1,2..}G^{n}(s)
\ee
s-proper time, and for $G^{n}(s)$ one can obtain:
\be
G^{n}(s)=\int{(Dz)^{\omega}_{on}exp(-K)\hat{tr}_{a}<W^{a}_{\Sigma}(C_{n})>}
\ee
where $K=\frac{1}{4}\int^{s}_{0} d\tau(\frac{dz^{\mu}}{d\tau})^{2}$, and $W^{a}_{\Sigma}(C_{n})$ is the adjoint  Wilson loop
defined for the gluon path $C_{n}$, which has both temporal (i4) and spacial projections (ij), and $\hat{tr}_{a}$  is the normalized
adjoint trace.   When $T>T_{c}$  the correlation function between CE and CM fields is rather week \cite{3}:
\be
<E_{i}(x)B_{k}(y)\Phi(x,y)\approx 0 
\ee  
and therefore, the expression for the Wilson loops is factorized  \cite{23}:
\be  <W^{a}_{\Sigma}(C_{n})>=L^{(n)}_{adj}(T)<W_{3}> \ee 
 with  $L^{(n)}_{adj} \approx L^{n}_{adj}$ for $T \le 1$ GeV.
One can integrate out  the $z_{4}$ part of the path integral $(Dz)^{\omega}_{on}=(Dz_{4})^{\omega}_{on}D^{3}z$, with the result 
\be G^{(n)}(s)=G^{(n)}_{4}(s)G_{3}(s)  ,  G^{n}_{4}(s)=\int (Dz_{4})^{\omega}_{on}e^{-K}L^{(n)}_{adj}=\frac{1}{2\sqrt{4\pi s}}e^{-\frac{n^{2}}{4T^{2}s}}L^{(n)}_{adj} \ee
This  factorization holds also  for quarks and will be used below (changing  the adjoint representation  for  the fundamental one).\par

The resulting gluon  contribution is 
\be
P_{gl}=\frac{N^{2}_{c}-1}{\sqrt{4\pi}}\int^{\infty}_{0}\frac{ds}{s^{3/2}}G_{3}(s)\sum_{n=0, 1, 2,...}e^{-\frac{n^{2}}{4T^{2}s}}L^{n}_{adj}, 
G_{3}(s)=\int (D^{3}z)_{xx}e^{-K_{3d}}<\hat{tr}_{a}W^{a}_{3}> \label{eqg}
\ee 
To  account for CMC    one can  introduce an  approximate expression for  3d Green function \cite{23}:
\be
G_{3}(s)=\frac{1}{(4\pi s)^{3/2}}\sqrt{\frac{(M^{2}_{adj})s}{sinh(M^{2}_{adj})s}},
M_{adj}\approx 2M_{D} \ee where $M_{D}$ is the gluon Debye mass. It should be mentioned that the eq.(\ref{eqg})  is in   a good agreement with the lattice data\cite{42*}.  \par
For quarks   one can write the expression of  the same form as  in (\ref{eqg}), but with the quark mass term $e^{-m^{2}_{f}s}$, and the density term $cosh(\frac{\mu n}{T})$
\be 
P_{f}=\sum_{q=u,d,s}P_{q},
P_{q}=\frac{4N_{c}}{\sqrt{4\pi}}\int^{\infty}_{0}\frac{ds}{s^{3/2}}e^{-m^{2}_{q}s}S_{3}(s)\sum_{n= 1, 2,...}(-)^{n+1}e^{-\frac{n^{2}}{4T^{2}s}}L^{n}_{f} cosh(\frac{\mu n}{T}) \label{eqf}\\
S_{3}(s)=\frac{1}{(4\pi s)^{3/2}}\sqrt{\frac{(M^{2}_{f})s}{sinh(M^{2}_{f})s}}, 
M^{2}_{adj}=\frac{9}{4}M^{2}_{f},L^{f}_{n}=(L^{adj}_{n})^{4/9} \ee
 The  full pressure reads as:
\be
P_{tot}=P_{f}+P_{gl}
\ee
Integrating over  ds  in (\ref{eqf})  one  obtains:
 \be P_{f}=\sum_{q=u,d,s}P_{q},\frac{P_{q} (T,\mu)}{T^4} = \frac{2 N_c}{\pi^2} \sum_n \frac{(-)^{n+1}}{n^2} \cosh \left( \frac{\mu n}{T} \right) L^n K_2  \left( \frac{\bar M n }{T}\right) \frac{\bar M^2}{T^2}, \label{press}\ee 
 where $\bar M =  \sqrt{m^2_f + \frac{M^2(T)}{4}}, ~~ M(T) = b\sqrt{\sigma_s (T)} $, b-is  the coefficient,   defined in \cite{23,25}.
 
 The sum (\ref{press}) can be  brought to the form 
 \be \frac{P_q (T,\mu)}{T^4} = f_+ (T,\mu) + f_-(T,\mu),\label{3.8}\ee
 
 \be f_{\pm} (T,\mu) =\frac{N_c}{3\pi^2} \int^\infty_0 \frac{ dz\left(z^2+ 2z \frac{\bar M}{T}\right)^{3/2}}{1+\exp \left( z + \frac{\bar M}{T} +\frac{ V_1}{2T} \mp \frac{\mu}{T}\right)},\label{sing}\ee
 where we have taken into account that $L=\exp \left( - \frac{V_1(\infty, T)}{2T}\right)$. 
\par
Analytical study of Eqs (\ref{press}),(\ref{sing}) needs some efforts and was done in \cite{25}. Two limits are simply done,one is the Stefan-Boltzman limit at high T,and
our result is some 15-20 $\%$ below this limit mainly due to CMC effects,and  another is the free quark limit with M tends to $m_q$ and $V_{1}=0(L=1)$ at extremely low temperatures, at this conditions the Fermi sphere is forming .    \par
 The expression (\ref{3.8}) has no singularities at  real $\mu$, but $f^{\pm}$ may get a singularity for imaginary chemical potentials for $Im(\mu)=\pi T$
due to vanishing of the denominator in (\ref{sing}) at $z=0$.\par
So one can conclude that in the normal situation with real  $\mu$ and $L_{f}$  the singularity is absent, this conclusion implies that there is no critical point $T_{c}(\mu)$  and the analytic structure  is affected only by the complex singularities. From this point of view, it seems that our consideration  could be  extended without any changes to a  large enough  values of the chemical potential as well as    in the case of finite temperatures for $T\le 1$   GeV  (or even higher with some modifications of the theory),but as we will show in case of finite  and large ($\mu_q \approx 600$ MeV) chemical potential such a trivial extension becomes inconsistent.

	\section{Speed of sound  in  a dense matter}
	The  sound velocity  is an  important quantity for the  description of a medium. For example it plays the crucial role   in the cooling  process in heavy ion collisions(\cite{Hydro1,Hydro2,Rev}), or in physics   of neutron stars(see e.g the part 5.15 of the book \cite{bookHen}). \par

Bounds and restrictions on the speed of sound is a  compelling topic \cite{HIC8,speed1,speed2,speed3,speed4,speed5,speed6,speed7,speed8,speed9,speed10,speed11,speed12,speed13}. The bound $0< C^{2}_{s} \le 1$ follows from thermodynamic stability  and causality\cite{speed13,speed14}. The scale $C^{2}_{s}=\frac{1}{3}$ is a general property of conformal theories with  the vanishing trace of the energy-momentum tensor $\epsilon-3P=0$. In the limit of very high temperature or very large density  QCD becomes nearly scale invariant  and  one  expects that $C^{2}_{s}\sim \frac13$. To describe  the scale invariance breaking  one can introduce two parallel quantities: a) the conformality measure $\Delta \nu^{2}_{s}=\frac13-C^{2}_{s}$ \cite{speed8}, and b) the interaction measure,or trace anomaly\cite{latt3,lattglu}.   \par
In case of finite temperatures  at $\mu=0$  the speed of sound is defined as:
\be C^{2}_{s}  =\frac{s}{\frac{\partial \varepsilon}{\partial T}} = \frac{\frac{\partial P}{\partial T}}{T\frac{ \partial^2 P}{\partial T^2}},~~ \mu=0.\label{3.1}\ee At large $T$  it  tends assymptotically to 1/3 \cite{latt3,qlatt2,qlatt1}. 
There are several possibilities to define the s.v  
at nonzero $\mu$ \cite{svdef}, in this paper we will focus on the isoentropic definition  i.e $s/n =$ const:
\be C^2_s = \frac{ n^2\frac{\partial^2 P}{\partial T^2} - 2sn \frac{\partial^2 P}{\partial T\partial \mu}+  s^2 \frac{\partial^2 P}{\partial \mu^2}}{(\varepsilon +p) \left(\frac{\partial^2 P}{\partial T^2}\frac{\partial^2 P}{\partial \mu^2}- \left( \frac{\partial^2 P}{\partial T\partial\mu}\right)^2\right)}=\frac{1}{\kappa_{s}(\epsilon+p)},\label{3.2}\ee
where we have defined:
\be s=\frac{\partial  P}{\partial T },~~ n=\frac{\partial  P}{\partial \mu},~~\varepsilon +P = Ts +\mu n. \label{3.3}\ee 
and the adiabatic compressibility 
\be \kappa_{s}=\frac{\frac{\partial^{2}P}{\partial T^{2}}\frac{\partial^{2}P}{\partial\mu^{2}}-(\frac{\partial^{2}P}{\partial T \partial \mu})^{2}}{n^{2}\frac{\partial^{2}P}{\partial T^{2}}-2sn\frac{\partial^{2}P}{\partial T \partial \mu}+s^{2}\frac{\partial^{2}P}{\partial\mu^{2}}} \ee

One can easily find that in  the limit of  zero $\mu$  one has 
 $ \frac{\partial  P}{\partial \mu}=0,~~\frac{\partial^2 P}{\partial T\partial \mu}=0 ,$\\ $ \frac{\partial^2 P}{\partial \mu^2}\neq 0$,  with the result: 
 \be C^2_s (\mu \to 0) = \frac{s^2}{( \varepsilon  +P)  \frac{\partial^2 P}{\partial T^2} }= \frac{   \frac{\partial  P}{\partial T }}{T    \frac{\partial^2 P}{\partial T^2}}, \label{3.4}\ee
 which coincides with  (\ref{3.1}).
 
 In the opposite limit, when $\mu$ is  larger than any scaleful  quantity, e.g. quark masses, screening masses of quark and gluons, one can write 
 \be P=T^4 f (\mu/T).\label{3.5}\ee
 
 Inserting (\ref{3.5}) into (\ref{3.2}), one has $\varepsilon +P = 4 f T^4$ and finally one  obtains the limit 
 \be C^2_s =\frac{4f(4ff''-3f'^{2})}{12f (4 ff''-3f'^{2})}=\frac13.\label{3.6}\ee
 
 As a next step we  calculate  the s.v for  the pressure given in (\ref{press}).
 One should note that the ratio $\frac{ \bar M}{T} $ at small quark mass $m_f$ has only logarithmic dependence in $T, \frac{\bar M(T)}{T} = b c_\sigma g^2 (T)$,  which can be neglected at not large $T, T\sim T_c \div 4 T_c$.
 
 Considering now large $\mu/T \gg 1$, one can neglect $f_- (T, \mu)$ and writing $f_+ (T,\mu) = f\left( \frac{ \mu- \frac{V_1}{2}}{T}\right)$, one obtains the following form \footnote{At this moment we are neglecting the dependence of $V_{1}$ on the chemical potential}
 \be C^2_s = \frac{4f^2 f'' - 3f'^2 f - \frac14 f'^3\Delta}{3f \left( 4 f f'' - 3 f'^2 - \frac{f'' f'}{3} \Delta\right)}, ~~ \Delta = \frac{T}{2} \frac{    \partial^2 V_1}{\partial T^2}.\label{3.10}\ee
 
 The form (\ref{3.10})  contains both the conformal limit when $\Delta \to 0$ and $C^2_s \to 1/3$ and the danger of strong derivations and possible singularities when $\Delta$ is large and positive (note that $4ff'' -3f'^2$ is positive).
 
 To clarify the matter we shall consider the case of  an arbitrary $\mu$, keeping the condition $\frac{\bar M}{T} =$const.
 
 Defining $x_\pm = \pm  \frac{\mu}{T} + b(T); ~~ b(T) =-\frac{\bar M}{T} - \frac{V_1(T)}{2T}$, one can write $f(x) = \int^\infty_0 \frac{ dz (z^2 +2z\bar M/T)^{3/2}}{1+\exp (z-x)};$
 $$ f_+ (T,\mu) = f_+ (x_+ (T)), ~~ f_-(T,\mu)=f_-(x_-(T)).$$
 
 $$ f\equiv f_+ (x_+) + f_- (x_-), ~~ f'_+ =\frac{\partial f_+}{\partial x_+},  {\rm etc.}, ~x'=\frac{dx}{dT}.$$

 The resulting expression for s.v. acquires the form 
 \be 
 C^2_s =\frac13 \frac{f^2 f'' - \frac34 f (f'_+-f'_-)^2 + \frac{A}{16} (f'_+-f'_-)^2+ B}{\left( 1+ \frac{Tb'f'}{4f}\right) \left( f^2f'' - \frac34 f (f'_+-f'_-)^2 + \frac{fC}{12}\right)}, \label{3.11}\ee
where we have defined
\be A= 2T (f'_+x'_+ + f'_-x'_-) +  
T^2 (f'_+x''_+ + f'_-x''_-),\label{3.12}\ee

\be B= (f''_+f'^2_-+   f''_-f'^2_+) \frac{b'^2}{4} + \frac12 f(f''_-f'_+ + f''_+f'_-) (x'_++x'_-),\label{3.13}\ee
\be C= 4b'^2 f_+'' f''_- + (2b'+b''T) (f''_+f'+f''_-f')+ 12b' (f''_+f'_- + f''_- f'_+).\label{3.14}\ee

One can again consider large $\mu$, when $f_-\ll f_+$ and then  $C\simeq (2b'+b''T) f''_+ f'_+) =- \frac{V''_1}{2T} f''_+ f'_-$, and $\frac{fC}{12} <0$ in the  denominator of (\ref{3.11}). One can estimate the numerical value of $\Delta = \frac{T}{2} \frac{\partial^2V_1}{\partial T^2}$ for  the Polyakov lines in the lattice data, and e.g. for the data of Bazavov et al. \cite{Baz} one finds that $\Delta (T)$ in the  interval  $T= (170\div 400)$ MeV  is changing  approximately  from 5 to 0.4 which shows a danger of $C \sim - \Delta (T)$ to cancel in the denominator, producing high and possibly negative values of $C^2_s$ , which may indicate instability. \par
Another interesting  preliminary conclusion is  that the sound velocity could exceed   $1/\sqrt{3}$,  both  properties should be  checked numerically, as it is done in the next section. 

\section{Numerical results}

\begin{figure}[h!]
				\centering
{\includegraphics[scale=0.2]{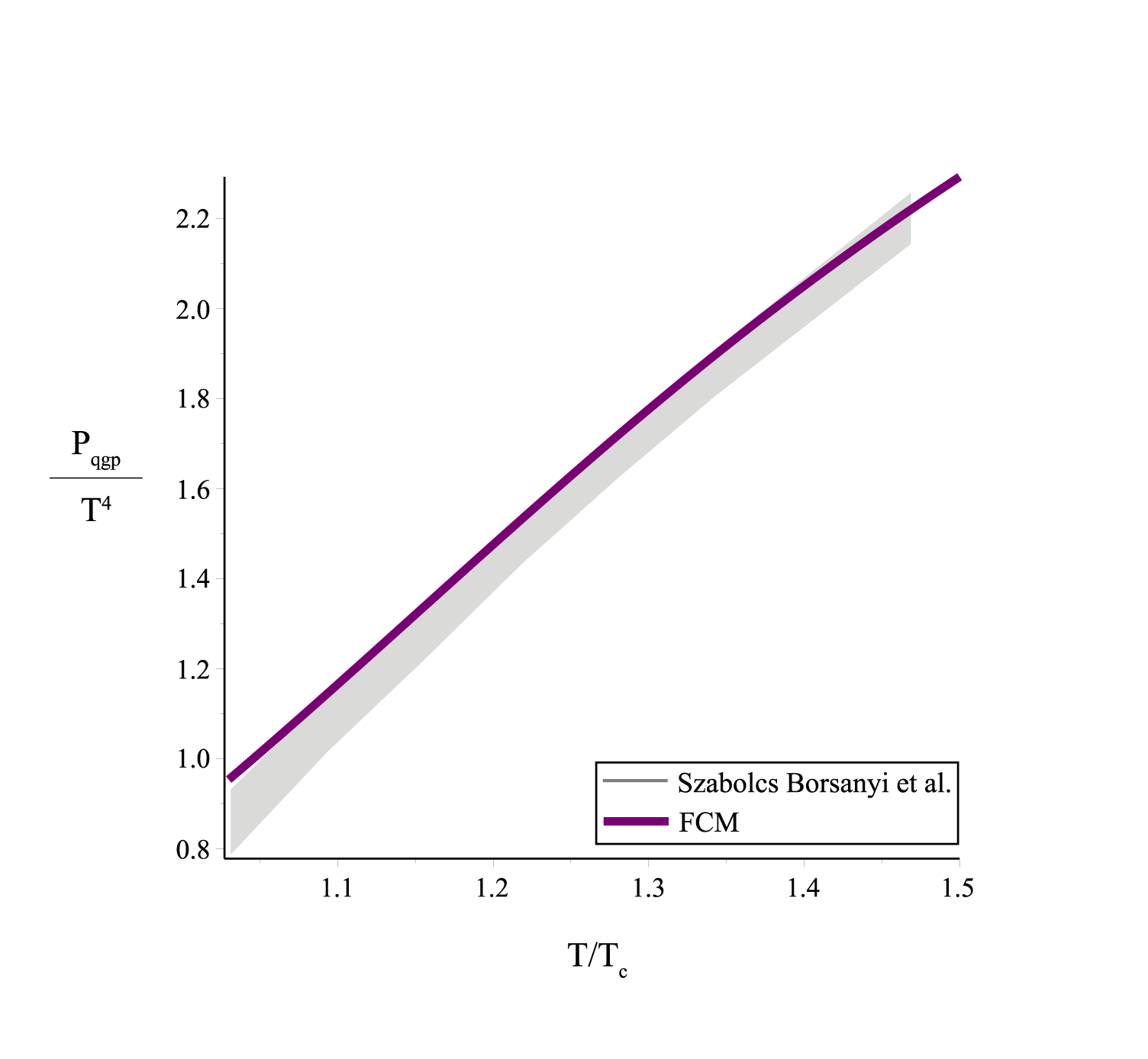}} 
									\caption{The pressure in QGP from FCM for  $\mu_{B}=0$ MeV, in comparison with the lattice data of Borsanyi et al.  \cite{thLatt2}    } \label{FIG1}
		\end{figure}

\begin{figure}[h!]
				\centering
{\includegraphics[scale=0.2]{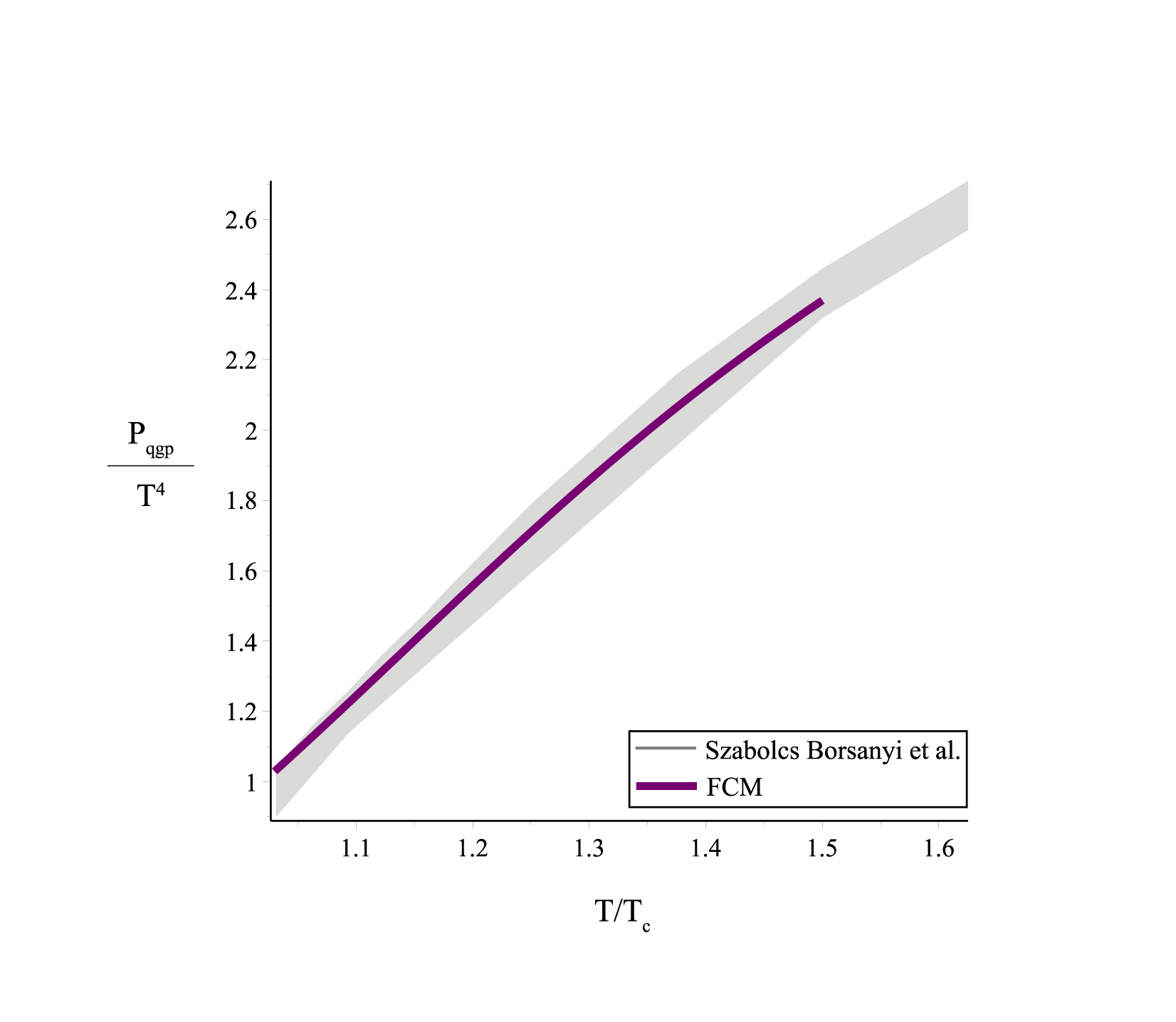}} 
									\caption{The pressure in QGP from FCM for $\mu_{B}=200$ MeV, in comparison with the lattice data of Borsanyi et al.  \cite{thLatt2}   } \label{FIG2}
		\end{figure}

\begin{figure}[h!]
					\centering
{\includegraphics[scale=0.2]{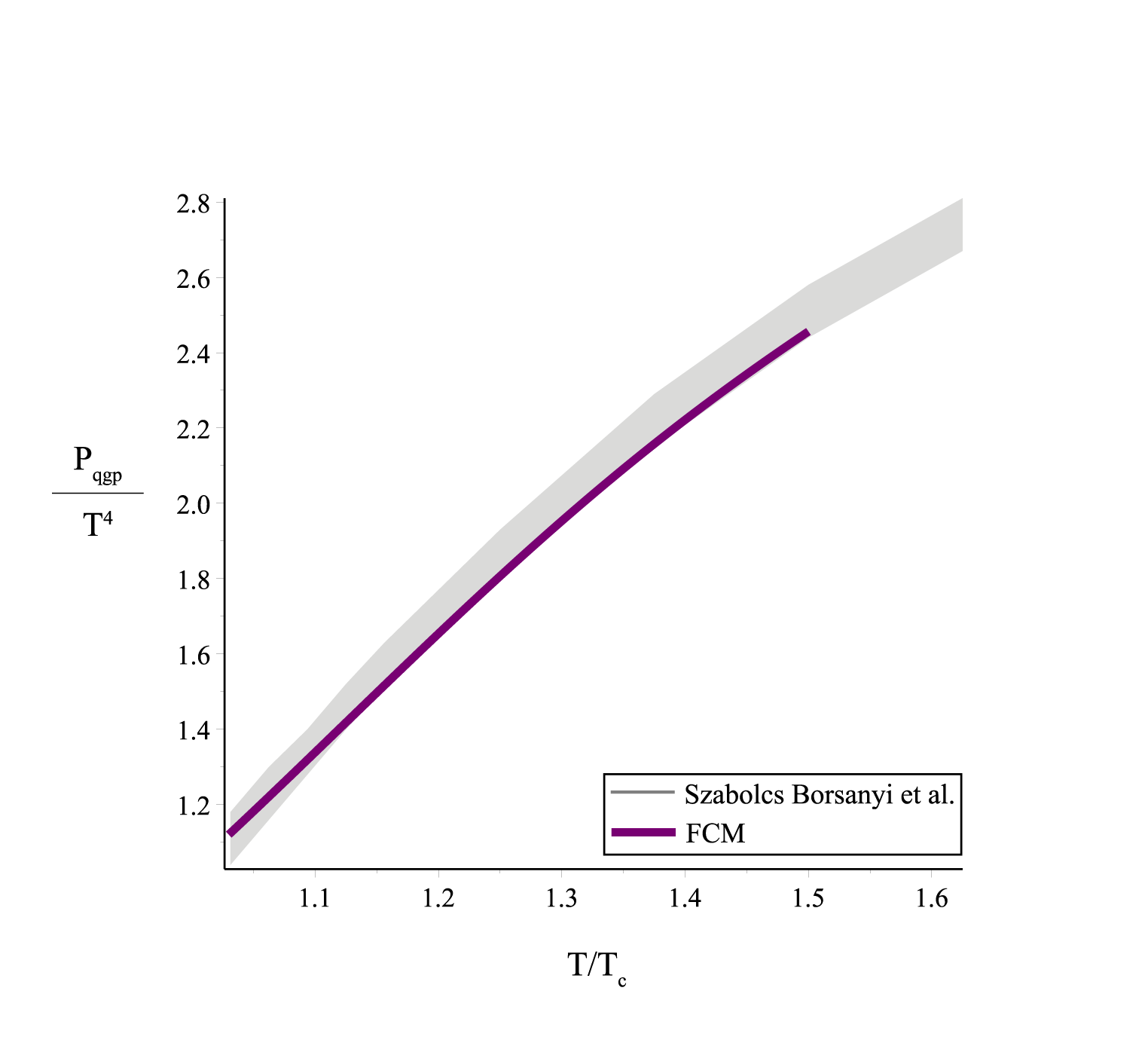}} 
									\caption{The pressure in QGP from FCM for $\mu_{B}=300$ MeV, in comparison with the lattice data of Borsanyi et al.  \cite{thLatt2}   } \label{FIG3}
		\end{figure}

\begin{figure}[h!]
	\centering
{\includegraphics[scale=0.2]{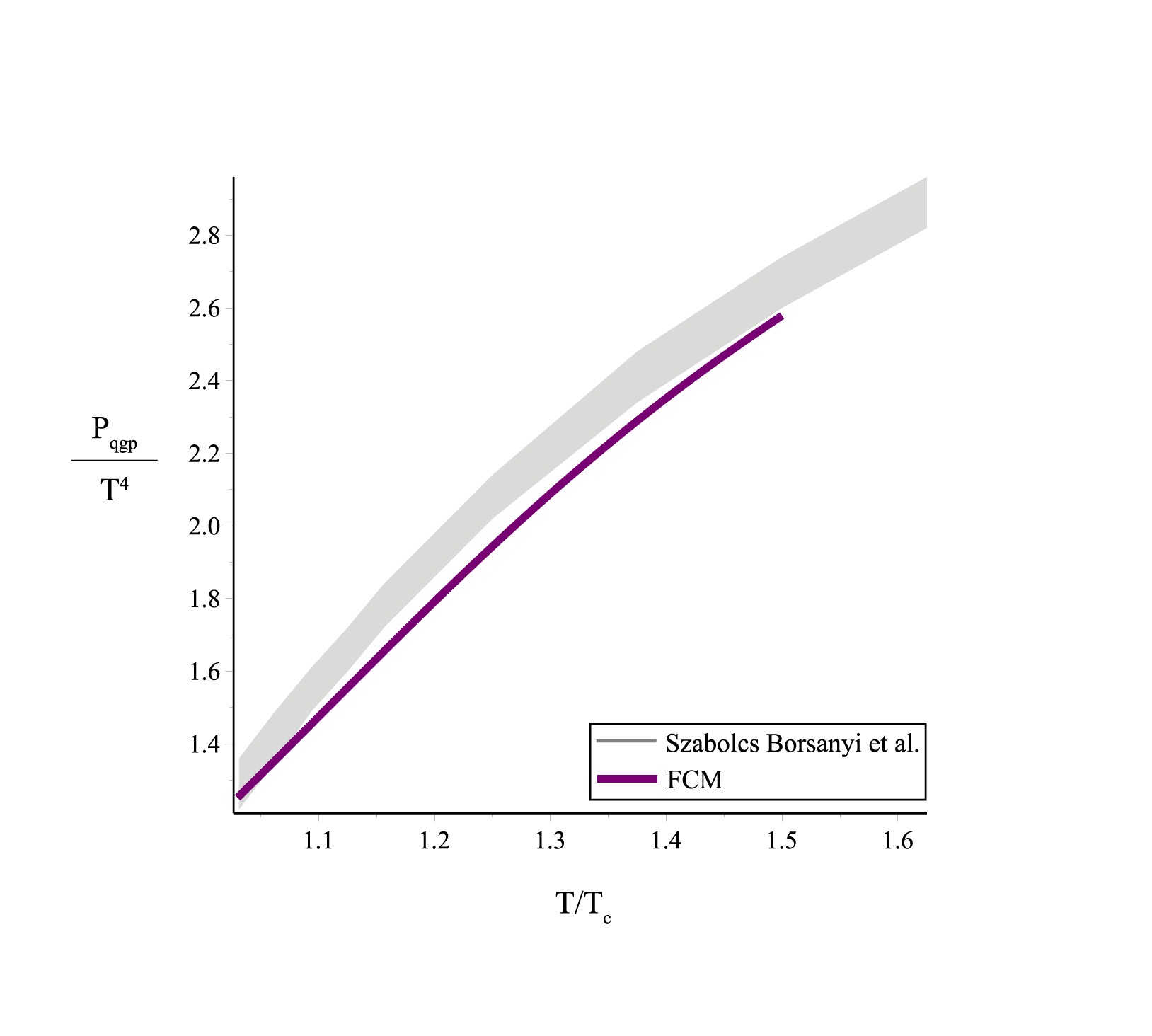}} 
									\caption{The pressure in QGP from FCM for  $\mu_{B}=400$ MeV, in comparison the lattice data of Borsanyi et al.  \cite{thLatt2} } \label{FIG4}
		\end{figure}

\begin{figure}[h!]
				\centering
{\includegraphics[scale=0.27]{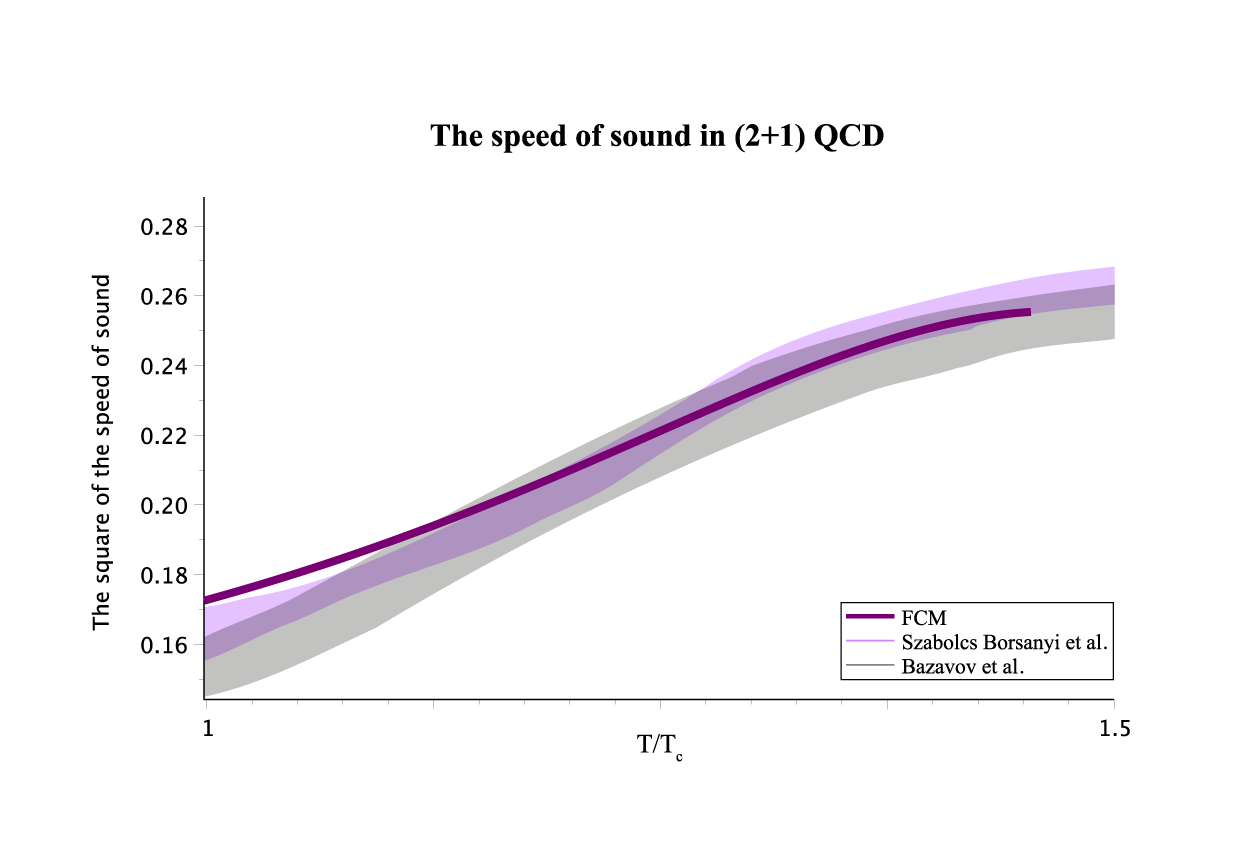}} 
									\caption{The square of the speed of sound in QGP from FCM(\ref{3.2}) for  $\mu_{B}=0$ MeV, in comparison with the data of Borsanyi\cite{thLatt2} et al. and Bazavov et al.  \cite{latt3}    } \label{FIG5}
		\end{figure}

\begin{figure}
		{\includegraphics[scale=0.27]{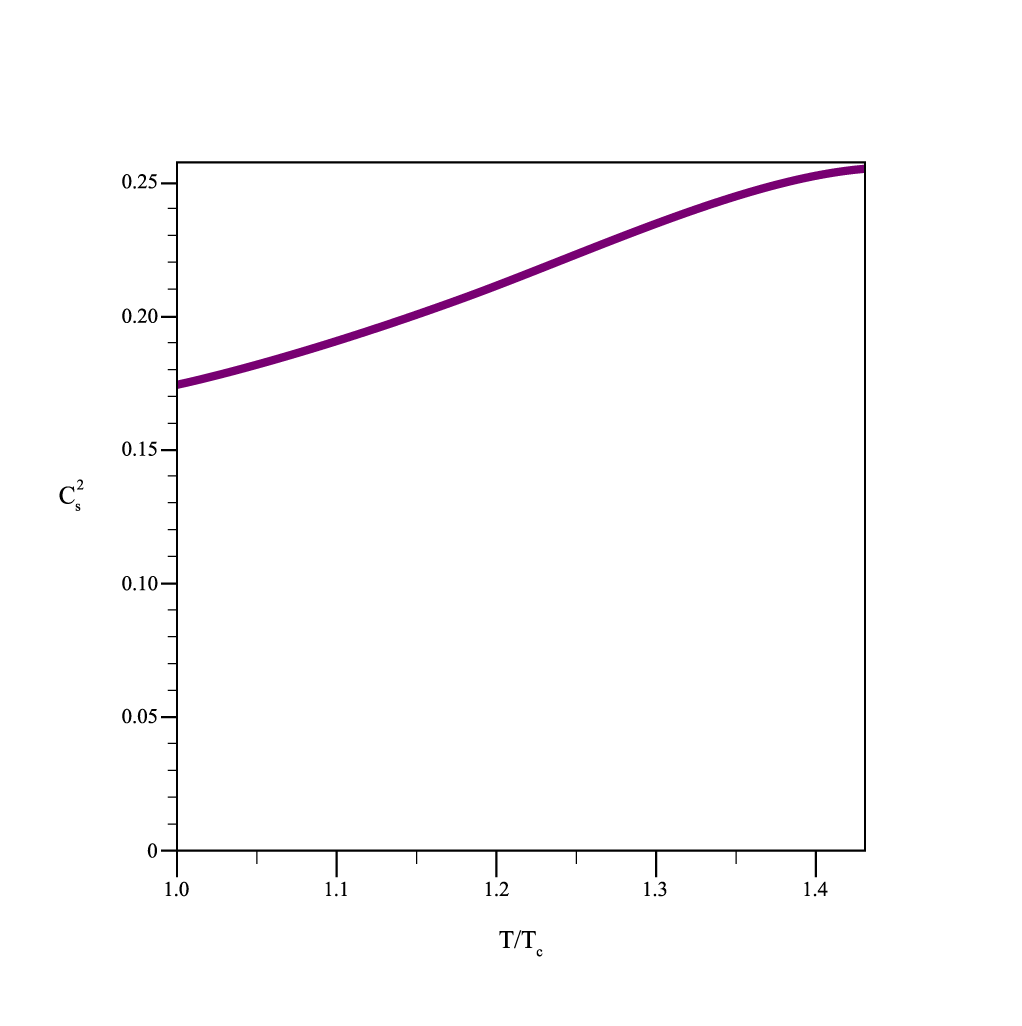}} 
									\caption{The speed of sound $C^{2}_{s}$ from FCM  (\ref{3.2}) for $\mu_{B}=0..400$MeV, the width of the line  that is shown  is equal to the change in the speed of sound in  this range       } \label{FIG6}
		\end{figure}

We will investigate  the behaviour of the speed of sound  at  temperatures from 160 Mev till 235 MeV. 
 According  to  the predictions of  the previous section    one can expect some  problems with the speed of sound  in this interval.
First of all we   compare our results for the pressure  at small $\mu$ with the  lattice predictions   \cite{thLatt2} to check the initial  data before  turning to  larger $\mu$.
As one can see in figures (Figs.\ref{FIG1},\ref{FIG2},\ref{FIG3},\ref{FIG4},\ref{FIG5}\footnote{The  speed of sound in FIG.5 a little bit different from our previous work \cite{ourspeed},where another form of $L(T)$ was used. That was done to be sure that our results are not connected 
to the specific form of $V_{1}(\infty,T)$ in \cite{ourspeed} in the confinement region.We 
have checked that the instability also persists for L(T) from \cite{ourspeed}  }), our results are   close to the lattice data.\footnote{We  are using  below  some approximation $L_{ap} (T, \mu=0)$,$L(\mu,T)\approx \sqrt{L_{Baz}}  $ taken from [53]  for the  Polyakov line  $L(T, \mu=0)$, which lies between the corresponding lattice  curves in \cite{Baz}  and  \cite{BorPol}. The preliminary FCM data for $L(T, \mu=0)$ generally agrees with $L_{ap} (T, \mu=0)$ and is now  in preparation.} We also calculate   the speed of sound in the range   $\mu_{B}=[0, 400]$ MeV , presented in Fig.\ref{FIG6}. 

One should  notice   that  the  results for the pressure in \cite{thLatt2} were  obtained in the first order of the  square of the chemical potential, however  corrections from higher orders     are not so important for low  densities.

		\begin{figure}[h!]
				\centering
{\includegraphics[scale=0.65]{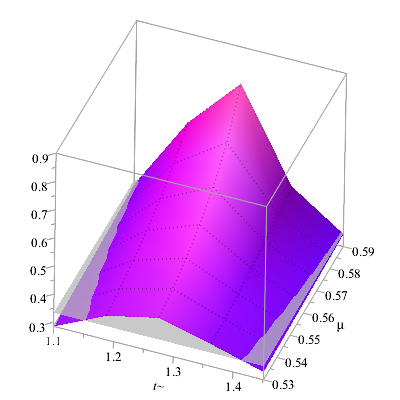}} 
								\caption{The square of the speed of sound in QGP from FCM. The semitransparent gray plane is at the value 1/3 .  } \label{FIG7}
		\end{figure}
		\begin{figure}[h!]
				\centering
{\includegraphics[scale=0.27]{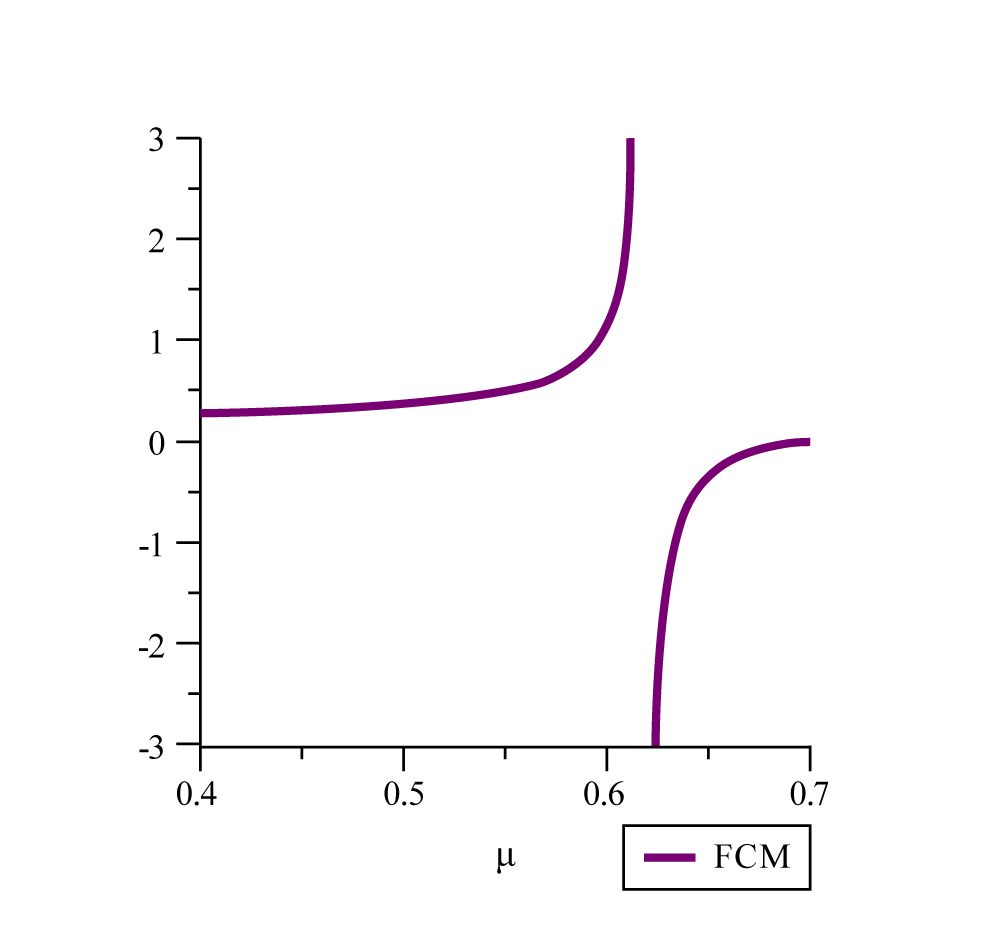}} 
									\caption{The square of the speed of sound in QGP from FCM at T=1.25 Tc. } \label{FIG10}
		\end{figure}
				\begin{figure}[h!]
				\centering
{\includegraphics[scale=0.65]{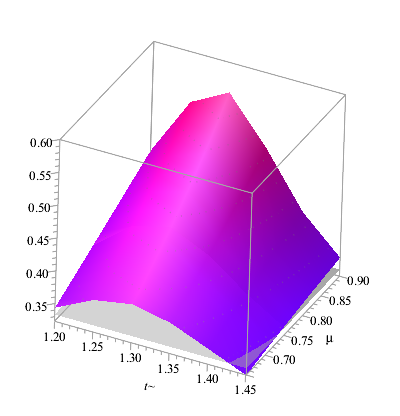}} 
									\caption{The square of the speed of sound in QGP from FCM with modified Polyakov Line as a function of T/$T_c$(left axis) and $\mu$ (right axis) in GeV. The parameter $a$=0.6.  } \label{FIG9}
		\end{figure}
			\begin{figure}[h!]
				\centering
{\includegraphics[scale=0.27]{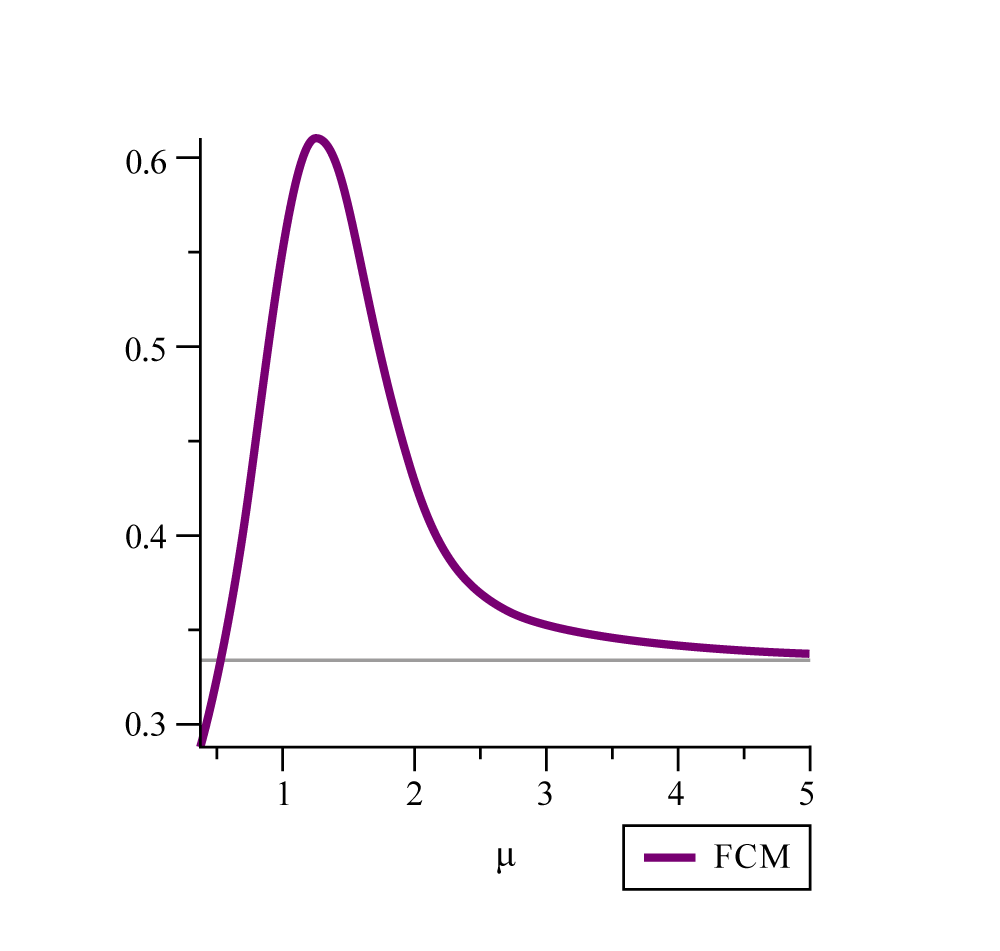}} 
									\caption{The square of the speed of sound in QGP from FCM with modified Polyakov Line as a function of $\mu$  in GeV. T=200 MeV. The parameter $a$=0.6.  } \label{FIG12}
		\end{figure}
		\begin{figure}[h!]
				\centering
{\includegraphics[scale=0.27]{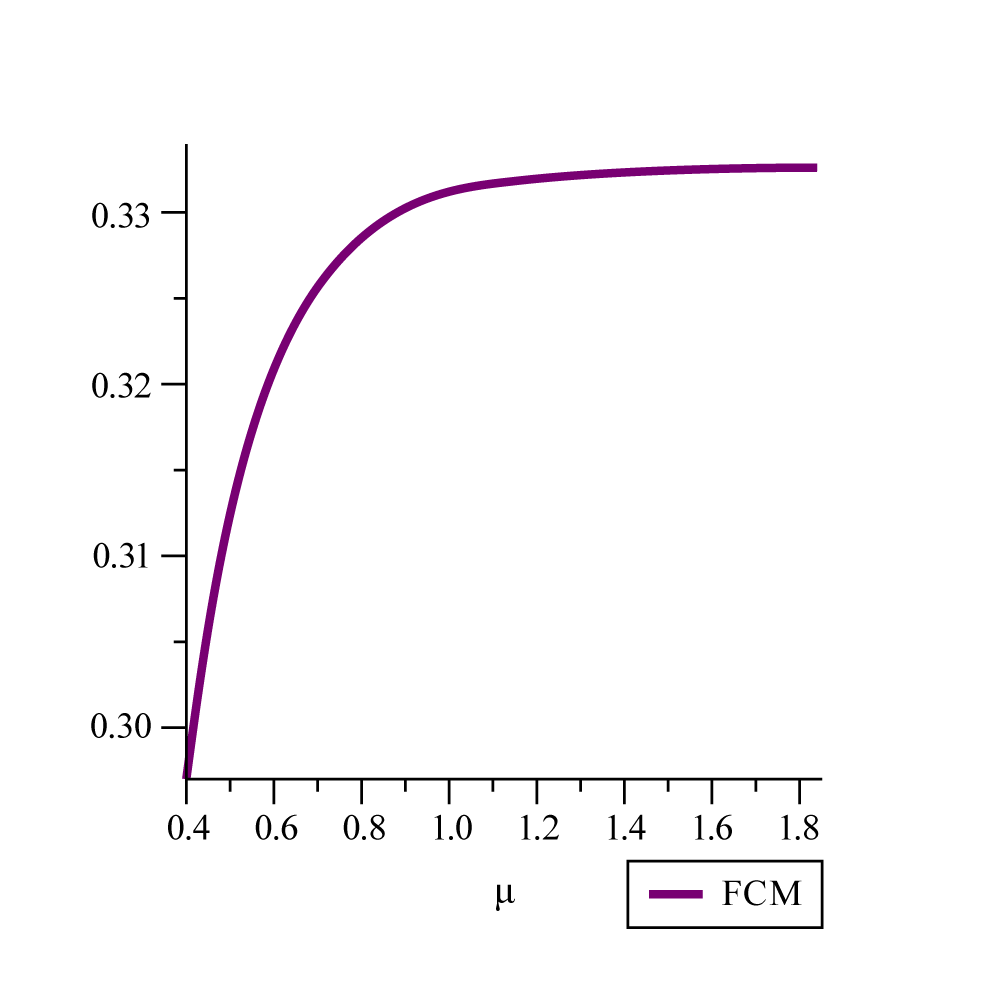}} 
									\caption{The square of the speed of sound in QGP from FCM with modified Polyakov Line as a function of $\mu$  in GeV. T=200 MeV. The parameter $a$=15.  } \label{FIG13}
		\end{figure}

\par  
 
We  plot  in Fig.\ref{FIG6} the changing in  the  speed of sound in the range $\mu_{B}=[0,400] $ MeV, where  the width of the line is equal to  the difference  $C^{2}_{s}(\mu_{B}=400)-C^{2}_{s}(\mu_{B}=0)$ .  One can see that at this densities there is no signal  about any  instabilities in the QGP.

\par

At high densities,   the speed of sound is found  from   the full expression   of the pressure (\ref{3.8}), (\ref{sing}) with inclusion of gluon contribution.

		As can be seen in Fig. \ref{FIG7}   there   exists a domain   where the square of the  speed  of sound exceeds   $\frac{1}{3}$.We treat this result with caution because, as was mentioned above, we use the fit for   the Polyakov loop  obtained for $\mu=0$. We also  find  domains  in Fig. \ref{FIG10} where the speed of sound  becomes negative. Obviously, because of the continuity of the denominator of (\ref{3.2}) as a function of $\mu,T$     the s.v. should exceed   the speed of light before  it  becomes negative.

		These results   demonstrate  a big difference between the finite temperature physics and  the physics of finite baryon densities. In the first case there is no problems with the system behaviour and the speed of sound  has the upper limit ($C^{2}_{2}\le \frac 13   
	$)   that it never exceeds. At finite temperatures ($0.16$ GeV $<T<1$ GeV) and zero $\mu$ the FCM describes physics of QGP(see for example \cite{ourspeed}. In the second case the system is  safe in the domain of  low chemical potentials and in the domain of very high chemical potentials but in the intermediate range its behaviour displays singular features,signalling the instability
of the thermodynamics in the QGP phase,which might mean either the emergence
of a new phase with a new dynamics or a necessity to correct the assumed 
dynamics,e.g. by taking the mu-modified Polyakov lines. 
 \par
 
  In the next section we will discuss possible corrections to the Polyakov loop due to the presence of the finite chemical potential,and we will see that  these corrections may change the picture of the instability.

	\section{Polyakov lines at nonzero $\mu$}

We investigate below how the nonzero baryon density can influence the  magnitude of Polyakov lines in the $qgp$.

To this end  we  write the pressure for  the quarks of the flavor $f$ in the np vacuum of $qgp$, where are present vacuum  averages of the colorelectric (CE) interaction $V_1 (r,T)$ and  colormagnetic (CM) interaction,  producing the screening mass $M(T)$, (see \cite{25} for details).

\be \frac{1}{T^4} P^{(f)}_q = \frac{N_c}{4\pi^2} \sum^\infty_{n=1} \frac{(-1)^{n+1}}{n^4} L^n (T) \cosh \frac{\mu n}{T} \phi_n (T)\label{4.1}\ee
and  $\phi_n(T)$ is given by \cite{25}
\be \phi_n(T) = \frac{8 n^2 \bar M^2}{T^2} K_2 \left( \frac{\bar M n}{T}\right), ~~\bar M = \sqrt{m^2_f + \frac{m^2(T)}{4}},\label{4.2}\ee
and $K_2$ is a modified Bessel  function.

Here $L(T)=$ is expressed via $V_1(r,T)$ as for the isolated quark line, i.e. distant from other quark or antiquark lines, $V_1 (r,T) \to V_1(\infty, T)$, 
\be L(T) = \exp \left( - \frac{V_1 (\infty, T)}{2T} \right).\label{4.3}\ee
 One should have in mind, that the actual origin of the Polyakov line in the pressure is the vacuum average of the phase factor $\Lambda (C) = \exp (ig \int_c A_\mu dz_\mu)$ along the path of the quark. Indeed, using the Fock-Schwinger representation for the free energy of an isolated quark one has \cite{3,23}, neglecting CM interaction 
 \be \frac{1}{T} F^q_0 =- \frac12 S p \int ^\infty_0 \xi (s) \frac{ds}{s} e^{- sm^2_q + s (D^2_\mu - g F_{\mu\nu} \sigma_{\mu\nu})}\label{4.4}\ee
 and neglecting the spin-dependent term $\sigma_{\mu\nu} F_{\mu\nu}$ one has 
 \be \frac{1}{T} F^q_0 =- \frac12   \int ^\infty_0 \xi (s) \frac{ds}{s} d^4x \sum_n (Dz)^w_{xx}  e^{- K-sm^2_q} \lan \Lambda(C_n)\ran,  \label{4.5}\ee
 where the kinematic factor $K$ and the winding path measure $(Dz)^{w}_{xy}$ are defined in \cite{3,23}.
 
 As  a result one obtains  the vacuum averaged factor with a still undefined contour $C$
 \be \bar \Lambda(T) = \frac{1}{N_c} \lan  tr \Lambda (C)\ran. \label{4.6}\ee
 
 At this point one can introduce two important simplifications, which lead to the final result (\ref{4.3}).
 a) $\Lambda(C_n)$ is a phase factor along the quark trajectory $C_n$ from the poin $x_4=0$ to the gauge equivalent point $x_4 =n/T$. Introducing the colormagnetic confinement (CMC)  acting inside the closed loop, one can close the trajectory $C_n$ by a  straight line from $x_4 =n/T$ to $x\_4 =0$, and writing the same straight line in the opposite direction. In this way after vacuum averaging one obtains
 \be \bar \Lambda (T) = L_n (T) \exp (-\sigma_s \cdot \Sigma_2),\label{4.7}\ee
 where $\Sigma_2$ is the area in the 3d projection of the closed loop $C_n$, which produces finally the screening mass $M(T) $ in (\ref{4.7}), and $L_n(T)$ is the vacuum average of the integral along $z_4$ axis
 \be L_n (T)= \lan P\exp (ig \int^{n/T}_0 A_4 dz_4)\ran \simeq\exp \left(-n\frac{V_1(\infty,T)}{2T}\right).\label{4.8}\ee

 To calculate $L(T)$ in the  dense $QGP$ one can use the property of the Polyakov lines studied in \cite{TVCM1, TVCM2}, namely that the static quark free energy $F_Q (T) = \frac{V_1(\infty,T)}{2}$ is  connected to the static quark-antiquark free energy $F_{Q\bar Q} (r\to \infty, T)=2 F_Q(T)$, or in our notations the identity $V_1  (r\to \infty, T) = 2 \frac{V_1(\infty, T)}{2}$. 
 
 In this way one can define the correlator of Polyakov lines 
 \be \lan \tilde{tr} L_{\vex} \tilde{tr}  L_{\vey}\ran = P (\vex-\vey), ~~ \tilde{tr} = \frac{1}{N_c} tr,\label{4.9}\ee
 which can be written as \cite{TVCM1}
 \be P(\vex-\vey) = \frac{1}{N^2_c} \exp \left(-\frac{V_1(r,T)}{T} \right) + \frac{N^2_c -1}{N^2_c} \exp \left( (-\frac98 V_1(\infty) - \frac18 V_1 (r)\right)/T), \label{4.10}\ee
 where the proper renormalization is to be done, as in \cite{TVCM1}.

 Usually one exploits, as was mentioned above, $F_{Q\bar Q} (T) = V_1 (\infty, T)$ from $P(\infty)$ to define Polyakov line, but it is clear, that for high density the  smaller distances $|\vex -\vey|\equiv R$ between $Q$ and $\bar Q$  should  effectively enter the corresponding interaction $V_1 (R,T)$, so that    $ L(T) \to  L(T, \bar R)$.

 One can associate $\bar R$ at large $\mu$ with the quark density $n(\mu) =\frac{\partial P}{\partial \mu}$ as follows $\bar R = R_0 \left( \frac{ n(\mu)}{n(0)} \right)^{-1/3}$, where $R_0$ is  an average $ q\bar q$ distance  at $\mu=0$. 
  One can realize that $n(\mu)> n(0)$ and hence $\bar R (\mu) < \bar R (0)$. Since $V_1 (\bar R, T)$ satisfies the relation $\frac{\partial V_1 (R,T)}{\partial R} >0$ (see \cite{TVCM1} for details of behaviour of     $V_1 (\bar R, T)$, one  obtains the relation
  \be V_1 (\bar R (\mu), T) < V_1 (\bar R (0), T), L(T,\mu)> L(T,0).\label{39}\ee
  This  behaviour is  supported by  lattice data  for  $L(T,\mu)$ in the case of SU(2) \cite{Kot1,98}.
  To check this relation and to understand  the role of this effect on the  sound velocity we have calculated $C^2_s (\mu, T)$, assuming $V_1 (\infty, T, \mu)= \frac{V_1 (\infty, T)}{1+a\mu^2}$, with $a>0$.   We investigated properties of the speed of sound for different values of parameter $a$ , and obtained following results:\par
	The instability exists for $a \lessapprox 0.5$ \par
	
	At larger values the speed of sound never exceeds the speed of light but it could break the conformal  limit as it is shown in Fig.\ref{FIG12}, and tends to $\sqrt{1/3}$ from above. 
  
  The resulting curve is shown in Fig.\ref{FIG9}, and one can see, that $C^2_s$ approaches  the value $C^2_s \approx 0.55$ at $T/T_C =1.3$  and $\mu_q =0.9$ GeV. It is rather curious that there are values for the parameter, for which the square of the speed of sound does not exceed 1/3 Fig.\ref{FIG13}, and  tends to $1/3$ from below, in accordance with \cite{speed2,Sakai,D3D7,speedN2,speedholo} . Thus one can say that FCM could describe both cases with and without exceeding of the conformal limit.And which one of them is realized depend upon details of the microscopic theory,which should predict the explicit behavior
of the Polyakov line $L(T,\mu)$ as a function of $\mu$.

\section{Conclusions and discussions.}

 The present paper is devoted to the   effects of  baryon chemical potential $\mu$ in the  dynamics of QGP.
 
 It is an extension of the study of QCD thermodynamics at zero $\mu$ in terms of the sound velocity made in \cite{ourspeed},  and is in the line of the series of papers \cite{3,19,20,21,22,23,24,25,26} where the QCD thermodynamics   is worked  out on the basis of FCM. In  particular, it was shown in \cite{25}, that nonzero $\mu$ do not present any difficulty for the method up to $\mu_q =0.4$ GeV.
 
 In the present paper this topic was addressed from the point of view of sound velocity and in the  temperature range which corresponds to QGP (at  least  for small $\mu$). This   analysis with the help of the sound velocity allows to  find instability regions  independently of the visible $P(T)$ behavior, since s.v. is sensitive to the second derivatives of $P(T)$.
 
 As it is seen in Figs. \ref{FIG1},\ref{FIG2},\ref{FIG3},\ref{FIG4} which scan the region  $\mu_B =[0,400]$ MeV, there is a good agreement between the FCM results for the pressure $P(T)$ and the  lattice  data \cite{thLatt1}.   In this region of $\mu_B$ the sound velocity is rather insensitive to $\mu_B$, as it is  demonstrated in Fig.\ref{FIG6}  and $C^2_s$ is fully in the conformal domain. \
 
 However $\mu_B =400$ MeV corresponds to $\mu_q\approx 130$ MeV and the main analysis of the  present paper refers to the  region $\mu_q > 500$ MeV. In this region without  inclusion of the chemical potential in the Polyakov line  the  instability domains are discovered in Figs.\ref{FIG10}.
 
 One can see in Fig.\ref{FIG7}  a  spectacular area in the $(\mu_q, T)$ plane  $\mu_q\geq 0.5$ GeV and $T/T_c= 1.1 \div 1.4$, where $C^2_s$ exceeds the 1/3 limit and approaches the value 0.8. Even more dangerous  region is  shown in Fig.\ref{FIG10} for $\mu_{q}\geq 0.6$ GeV, where  $C^2_s$ may exceed unity and even become negative, which means the instability region, not subject to the  chosen dynamics.   
This results are obtained with the  Polyakov loop $L(T)$  independent of $\mu$. So one must expand a little on the importance of the $L(T)$ (and $L(T,\mu)$) dynamics. In FCM the Polyakov loop enters self-consistently   from  the  very beginning, since it  belongs to the quark and gluon propagators in the   formalism.

The importance of $L(T)$ was numerically proved in the  first stage of the FCM formalism in \cite{3,19,20,21}, where it was the main dynamical effect (without CMC).

In the present paper we have chosen in Figs.1-7  the Polyakov line values $L(T,\mu=0)= L_{ap} (T)$, which lie between the lattice data  for $L(T,0)$ obtained in \cite{Baz} and \cite{BorPol}, and is very close to the preliminary values of $L(T,0)$ obtained within the FCM formalism (to be published).

As was discussed  in section 5, the  introduction of  $\mu$ in $L(T)$ leads to the  decrease of $V_1(T)$ and  the  increasing of $L(T,\mu)$, and  as  it was discussed above in section 3, this leads to  smaller  values of $\Delta$ in  Eq. (\ref{3.10}), and $C$ in Eq. (\ref{3.14}),  i.e. the  terms which effectively lead to instabilities of $C^2_s$. To  prove  this   dependence and to investigate characteristics of the  instability(i.e.the effect of the singular behaviour of the sound 
velocity) we included the chemical potential in the Polyakov line in the following way $V=\frac{V_{1}(T)}{1+a\mu^2}$, and observed that the instability survives until  parameter $a$ exceeds critical value(in our case $a\approx 0.55$). With increasing of $a$-parameter  a dangerous region is shifted to larger $\mu$, and eventually the instability vanishes. 

When this happens the speed of sound becomes well behaved quantity but it could still exceed the speed of light, but for a $>0.58$ the last  problem is absent    \par
There is a range for $a$-parameter  where s.v breaks the conformal limit. Thats  clearly seen on    FiG.\ref{FIG9}  and FiG.\ref{FIG10}. In FiG.\ref{FIG10} the speed of sound as a function of chemical potential was shown. As one can see, in the limit of large potentials  the square of  s.v  tends to 1/3  from above, as predicted in (\ref{3.6}).With further increasing of the $a$-parameter, breaking of conformal limit becomes less significant and eventually it disappears FiG.\ref{FIG13}.

Summarizing one can say that  the dynamics of QGP is well described by the FCM formalism, which agrees with lattice data for $\mu=0$ and  predicts a reasonable extension to nonzero $\mu$, being however  parameter dependent.

\section{Acknowledgements}
The authors are grateful for useful discussions to M. A. 
Andreichikov and B. O. Kerbikov, M. A. Zubkov,E.A.Fedina,  R.A.Abramchuck and especially to S.I.Blinnikov for very fruitful discussions about physics of neutron stars.\par 
 This work was done in the frame of the scientific project, supported by the Russian Science Foundation grant number 16-12-10414.

\end{document}